# An Extensive Repot on the Efficiency of AIS-INMACA (A Novel Integrated MACA based Clonal Classifier for Protein Coding and Promoter Region Prediction)

Pokkuluri Kiran Sree[1], Inampudi Ramesh Babu[2]

[1] Research Scholar, Dept of CSE, JNTU Hyderabad
[2] Professor, Dept of CSE, Acharya Nagarjuna Univesity, Guntur

[1]profkiransree@gmail.com

*Abstract*

This paper exclusively reports the efficiency of AIS-INMACA. AIS-INMACA has created good impact on solving major problems in bioinformatics like protein region identification and promoter region prediction with less time (Pokkuluri Kiran Sree, 2014). This AIS-INMACA is now came with several variations (Pokkuluri Kiran Sree, 2014) towards projecting it as a tool in bioinformatics for solving many problems in bioinformatics. So this paper will be very much useful for so many researchers who are working in the domain of bioinformatics with cellular automata.

## Introduction

Protein coding regions of DNA(Datta, 2005) is an imperative segment of a cell and genes will be found in particular share of DNA which will hold the data as unequivocal arrangements of bases (A, G, C, T).these express successions of nucleotides will have guidelines to manufacture the proteins. Anyway the locale which will have the guidelines which is called as protein coding areas involves quite less space in a DNA arrangement. The ID of protein coding locales assumes an essential part in comprehension the gens.

Promoter DNA (Synder,2002) is an extremely significant part in a unit, which is found in the core. DNA holds part of data. For DNA succession to transcript and structure RNA which duplicates the obliged data, we require a promoter. So promoter assumes an essential part in DNA translation. It is characterized as "the grouping in the locale of the upstream of the transcriptional begin site (TSS)".

## Understanding of AIS-INMACA

TABLE 1 EXAMPLE DATASET

|     | Attribute 1 | Attribute 2 | Attribute 3 | Class |
|-----|-------------|-------------|-------------|-------|
| 1.  | 0.00 | 0.00 | 0.45 | C |
| 2.  | 0.75 | 1.00 | 0.00 | N |
| 3.  | 1.00 | 1.00 | 1.00 | C |
| 4.  | 0.00 | 0.25 | 0.45 | N |
| 5.  | 1.00 | 1.00 | 0.25 | C |
| 6.  | 0.50 | 0.75 | 1.00 | C |
| 7.  | 0.30 | 0.40 | 0.65 | C |
| 8.  | 1.00 | 1.00 | 0.75 | C |
| 9.  | 0.00 | 0.50 | 0.25 | C |
| 10. | 0.75 | 0.50 | 0.25 | C |
| 11. | 1.00 | 0.75 | 0.25 | N |
| 12. | 0.00 | 0.00 | 0.25 | N |
| 13. | 0.75 | 1.00 | 0.25 | N |
| 14. | 0.50 | 0.50 | 0.50 | C |
| 15. | 0.25 | 0.40 | 0.65 | N |
| 16. | 0.00 | 0.75 | 0.50 | C |
| 17. | 0.00 | 0.50 | 0.25 | N |
| 18. | 0.25 | 0.25 | 0.25 | N |
| 19. | 0.00 | 0.50 | 0.00 | C |
| 20. | 0.05 | 0.25 | 1.00 | N |
| 21. | 0.00 | 0.50 | 1.00 | C |
| 22. | 0.00 | 0.45 | 0.25 | N |

*AIS-INMACA (Non Complemented Rule)*

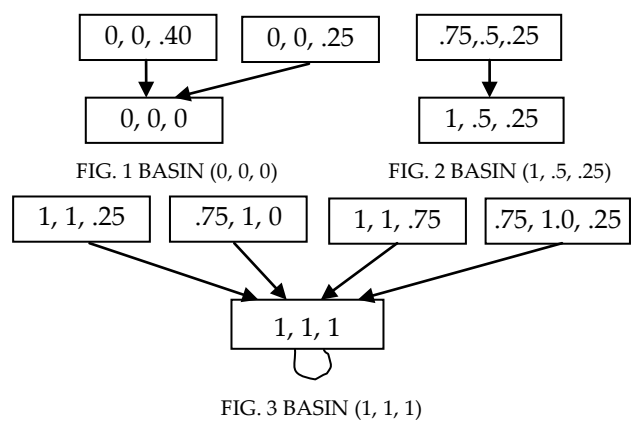

FIG. 1 BASIN (0, 0, 0)    FIG. 2 BASIN (1, .5, .25)

FIG. 3 BASIN (1, 1, 1)

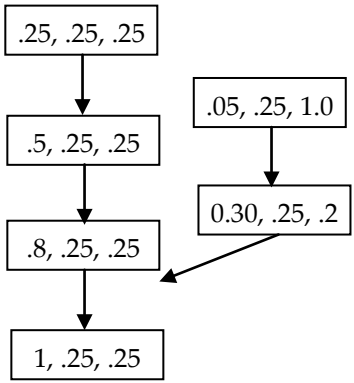

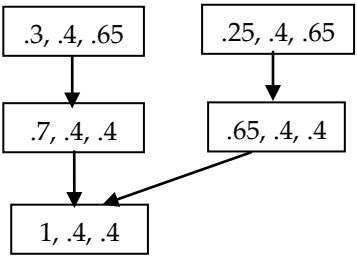

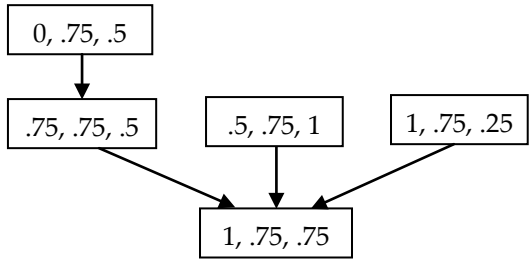

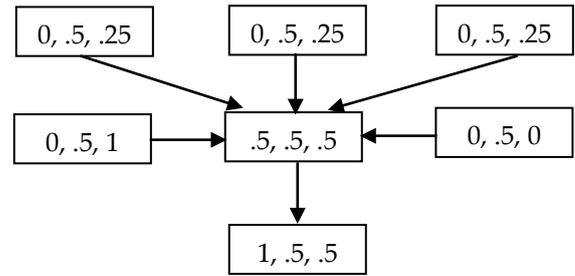

FIG. 7 BASIN (1, .5, .5)

### AIS-INMACA (Complemented Rule)

A Cellular Automata which uses fuzzy logic is an array of cells arranged in linear fashion evolving with time. Every cell of this array assumes a rational value in the interval of zero and one. All this cells changes their states according to the local evaluation function which is a function of its state and its neighboring states. The synchronous application of the local rules to all the cells of array will depict the global evolution.

Assume n represent the number of fuzzy states and qj denotes the fuzzy state, a rational value in between zero and one will assigned to each state.

$$Q_j = j/n-1 \text{ whre } j=1.1.2\ldots\ldots n-1.$$

Example if n=6, Cell can assume a rational values of $q_0=0, q_1=.20, q_2=.40, q_3=.60, q_4=.80, q_5=1$. Table 1 shows the example data set. Fig[1-7] shows the AIS-INMACA build up by non complemented rules. Fig[8-12] shows the AIS-INMACA build up by complemented rules.

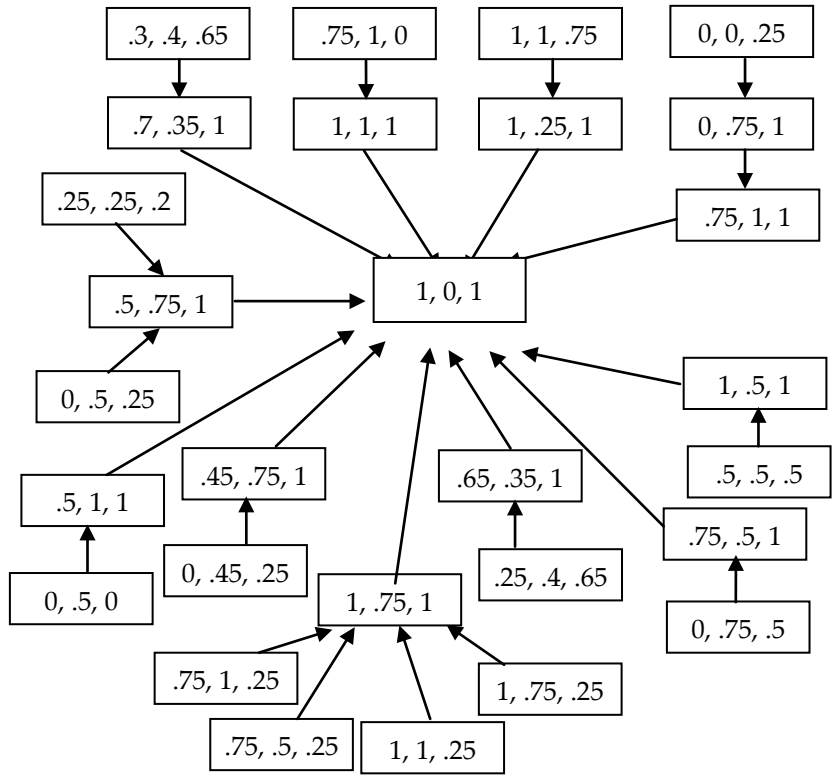

FIG. 8 BASIN (1, 0, 1)

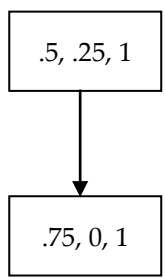
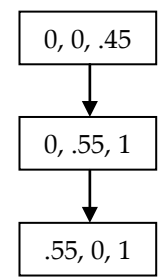

FIG. 9 BASIN (.75, 0, 1)    FIG. 10 BASIN (.55, 0, 1)

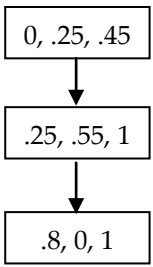
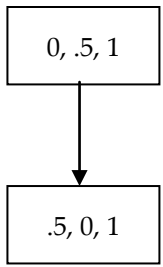

FIG. 11 BASIN (.8, 0, 1)    FIG. 12 BASIN (.5, 0, 1)

## Efficiency of AIS-INMACA

Burset and Guigo turned out with a paramount paper depicting strategies keeping in mind the end goal to test gene expectation programs. In this paper, they depict a set of known coding successions that ought to be utilized as information to prepare the models. Likewise, a set of known coding groupings is given to assess the achievement of the model.

The important statistics to look at include:

1. True Positives (TP): Number of correctly predicted coding regions
2. False Positives (FP): Number of incorrectly predicted coding regions
3. True Negatives (TN): Number of correctly predicted non-coding regions
4. False Negatives (FN): Number of incorrectly predicted non-coding regions

Using the above measures following are calculated

Actual Positives (AP) = TP + FN

Actual Negatives (AN) = TN + FP

Predicted Positives (PP) = TP + FP

Predicted Negatives (PN) = TN + FN

Sensitivity (SN) = TP/(TP + FN) (Percentage of coding regions found)

Specificity (SP) = TP/(TP + FP) (Percentage of positive that are correct)

These measures can be combined to form a single correlation coefficient:

$$CC = \frac{[(TP)(TN) - (FP)(FN)]}{\sqrt{[(AN)(PP)(AP)(PN)]}}$$

## Experimental Results

Examinations were led by utilizing Fickett and Toung information (Fickett, 1992) for foreseeing the protein coding locales. All the 21 measures reported in (Fickett, 1992) were acknowledged for creating the classifier. For promoter locale recognizable proof human promoters from Mpromdb(34,128) and Mamal promoter accumulation information sets from UCI Machine Learning Repository was taken for testing and preparing. Fig [13-17] shows the output of AIS-INMACA.

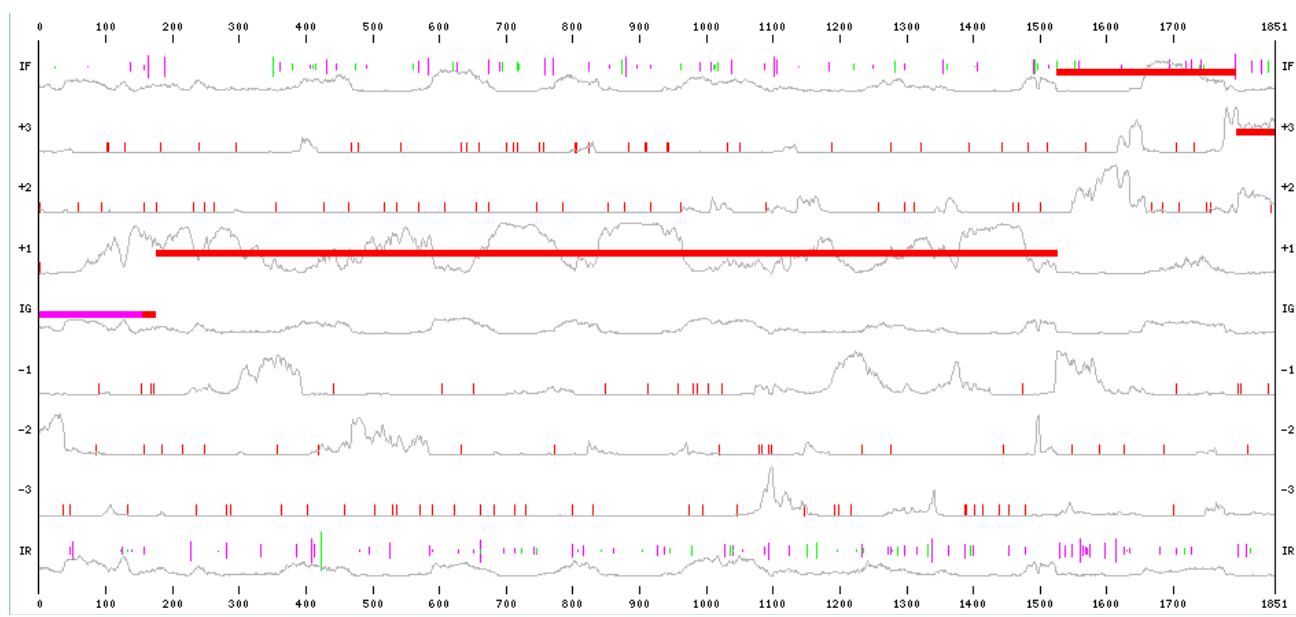

FIG. 13 EXONS FINDING

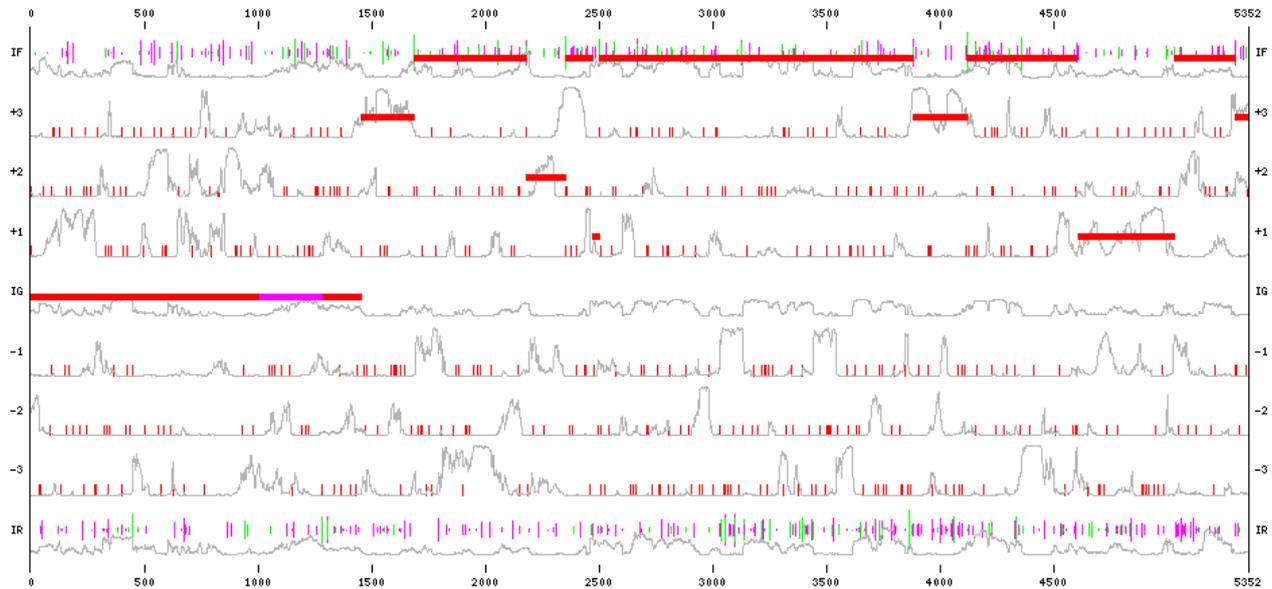

FIG. 14 BOUNDARIES OF EXONS

FIG. 15 EXONS FINDING

FIG. 16 BOUNDARY REPORTING

| Start | End | Score | Promoter Sequence |
|---|---|---|---|
| 20 | 70 | 0.94 | AGCCTATTTAGAAAAATGGGCCATTAGGAAATTGCAAGGAAGAACCATTC |
| 318 | 368 | 1.00 | GGGGGTGGGTATAAAAATGGGCAGTGCTCTGGGCCCTGTCACTGACGTTT |
| 1553 | 1603 | 0.81 | TGCAGGAATATTTAAATTCTCCTCCTATGCTTTGTCCCTAGCTTCCTCAT |
| 1704 | 1754 | 0.87 | TTTAGTATCCAAAAAAGGCAGTCCACAAAAGGTGATAGGAGATAAGCTGA |
| 4233 | 4283 | 0.88 | GTATCGCCGGCATAAGAGCCCCCTGCCTGAATAGCCACAAAGACTGGAGA |
| 4640 | 4690 | 0.91 | AGAGGAGCAGAATAAAGGGGGTTCCCTGGGGAAAGCATATCGCTTTTGTA |
| 5198 | 5248 | 0.85 | TTAACTTCTGTATATAAGTTCTGTTCTTATCCTACCAAAAAAAAAAAATC |
| 5287 | 5337 | 0.91 | TGTTTGCTCTATATCTATGCATGCGGGTGAGATGGCGGTGACTGCAGTGG |

FIG. 17 PROMOTER IDENTIFICATION

**Conclusion**

This paper proves AIS-INMACA is crucial in predicting both protein coding and promoter region identification. A number of variations like AIS-MACA-X and AIS-MACAA-Y are developed by us for improving the accuracy of classification extending the basic AIS-INMACA. This discussion projects AIS-INMACA as an important and useful classifier in bioinformatics. The average accuracy reported for protein coding prediction is 86% and promoter prediction is 87.6%.